\documentclass[
11pt,%
final,%
tightenlines,%
notitlepage,%
twoside,%
onecolumn,%
nobibnotes,%
nofootinbib,%
superscriptaddress,%
noshowpacs,%
noshowydk,%
showdoi,%
centertags,%
amsmath,
amsfonts,
maik]%
{revtex4-1}

\usepackage{graphicx}
\usepackage[colorlinks,citecolor=blue,urlcolor=blue]{hyperref}

\newcommand{\Eq}[1]{(\ref{eq:#1})}

\newcommand{\Fig}[1]{Fig.~\ref{fig:#1}}

\newcommand{\eps}{\varepsilon}

\newcommand{\txtQ}{{\text{Q}}}

\newcommand{\tr}[1]{\mathop{\rm tr}({#1})}

\newcommand{\beq}[1]{\begin{equation}\label{eq:#1}}
\newcommand{\eeq}{\end{equation}}

\newenvironment{se}[1]{\equation\label{eq:#1}\aligned}{\endaligned\endequation}
\newcommand{\bsplit}[1]{\begin{se}{#1}}
\newcommand{\esplit}{\end{se}}

\providecommand*{\twoD}{\textsc{2d}}
\providecommand*{\threeD}{\textsc{3d}}
\providecommand*{\fourD}{\textsc{4d}}

\newcommand{\Hen} {H\'enon}
\newcommand{\ahenon}{a_{\text{h}}}
\newcommand{\ahenOne}{a_{\text{h}1}}
\newcommand{\ahenTwo}{a_{\text{h}2}}

\newcommand{\ui}{\text{i}}

\newcommand{\bR}{{\mathbb{ R}}}

\newcommand{\movierefall}{For a rotating view see
\href{http://www.comp-phys.tu-dresden.de/supp/}{http://www.comp-phys.tu-dresden.de/supp/}.}

\begin{document}

%*************************************************************
~

\title{Moser's quadratic, symplectic map}
\author{\firstname{Arnd}~\surname{B\"acker}}
\affiliation{Technische Universit\"{a}t Dresden,
          Institut f\"{u}r Theoretische Physik and Center for Dynamics, 01062 Dresden, Germany}
\affiliation{Max-Planck-Institut f\"{u}r Physik komplexer Systeme, N\"{o}thnitzer Strasse 38, 01187 Dresden, Germany}
\author{\firstname{James~D.}~\surname{Meiss}}
\affiliation{University of Colorado, Department of Applied Mathematics, Boulder, CO 80309-0526, USA}

\begin{abstract}
In 1994, J\"urgen Moser generalized H\'enon's area-preserving quadratic map to obtain a
normal form for the family of four-dimensional, quadratic,
symplectic maps. This map has at most four isolated fixed points.
We show that the bounded dynamics of Moser's six parameter family is organized
by a codimension-three bifurcation, which we call a \textit{quadfurcation},
that can create all four fixed points from none.

The bounded dynamics is typically associated with Cantor families of invariant
tori around fixed points that are doubly elliptic. For Moser's map there can be two
such fixed points: this structure is not what one would expect from dynamics near
the cross product of a pair of uncoupled H\'enon maps, where there is
at most one doubly elliptic point.
We visualize the dynamics by escape time plots on \twoD{} planes
through the phase space and by \threeD{} slices through the tori.

~

\noindent
MSC2010 number: \texttt{37J40, 70H08, 34C28, 37C05}

~


\end{abstract}

%MSC Classification (Not PACS)
%\pacs{37J40, 70H08, 34C28, 37C05}
%       37J40   Perturbations, normal forms, small divisors, KAM theory, Arnold diffusion
%       70H08   Nearly integrable Hamiltonian systems, KAM theory
%       34C28   Complex behavior, chaotic systems
%       37C05   Smooth mappings and diffeomorphisms

\maketitle

%%%%%%%%%%%%%%%%%%%%%%%%%%%%%%%%%%%%%%%%%%%%%%%%%%%%%%%%%%%%%%%%%%%%%%%%%%%%%
\section{Introduction}
%%%%%%%%%%%%%%%%%%%%%%%%%%%%%%%%%%%%%%%%%%%%%%%%%%%%%%%%%%%%%%%%%%%%%%%%%%%%%

Moser derived a normal form for quadratic symplectic maps
on a $2n$-dimensional vector space in 1994 \cite{Moser94}.
For $n=1$ the normal form is conjugate to the
famous and well-studied H\'enon map \cite{Hen1976}.
In contrast, the dynamics of the Moser map for $n \ge 2$ have
not yet been explored. In \cite{BaeMei2018:p}
we started the investigation of the
$n=2$ case, i.e., the four-dimensional quadratic symplectic map.
Perhaps one of the most intriguing of its features is the possibility
of a \emph{quadfurcation}, in which four new solutions are
created out-of-nowhere under parameter variation.
Of particular interest are the possible combinations
of stability configurations; in particular, the case
when two of the fixed points are elliptic-elliptic and two are elliptic-hyperbolic
is surprising, as this would not be possible
when the different degrees of freedom are not coupled.

Symplectic maps naturally arise as Poincar\'e maps
of Hamiltonian flows and therefore are
relevant for many different applications,
for example in the dynamics of the solar system and galaxies
\cite{Wisdom91,MurHol2001,Cincotta02, Contopoulos13,PaeEft2015,DaqRosAleDelValRos2016},
beam dynamics of particle accelerators
\cite{Warnock92,Dumas93,RobSteLasNad2000,Pap2014},
plasma physics \cite{Howard86},
and chemistry
\cite{Gaspard89, Gillilan91,TodKomKonBerRic2005, Gekle06,WaaSchWig2008, ManKes2014}.
Investigating the simplest prototypical examples, such as the
area-preserving H\'enon map or Chirikov's standard map
in the two-dimensional case, helps to give
a good understanding of the way in which dynamical properties
depend on parameters in typical maps.
Thus the study of the dynamics of Moser's four-dimensional,
quadratic symplectic map is of fundamental interest
as it describes the generic local behavior
of symplectic maps in \fourD{}.

\newpage
%%%%%%%%%%%%%%%%%%%%%%%%%%%%%%%%%%%%%%%%%%%%%%%%%%%%%%%%%%%%%%%%%%%%%%%%%%%%%
\section{Normal form}
%%%%%%%%%%%%%%%%%%%%%%%%%%%%%%%%%%%%%%%%%%%%%%%%%%%%%%%%%%%%%%%%%%%%%%%%%%%%%

As Moser showed \cite{Moser94}, every quadratic symplectic map
on $\bR^{2n}$ can be written as the composition of an affine symplectic map,
a symplectic shear,
$\sigma: \bR^n \times \bR^n \to \ \bR^n \times \bR^n$, of the form
\beq{Shear}
	\sigma(x,y) = (x, y - \nabla V(x)) ,
\eeq
and a linear symplectic map. When the map is quadratic,
$V: \bR^n \to \bR$ is a cubic potential.

\begin{figure}[b]
\begin{center}
\includegraphics{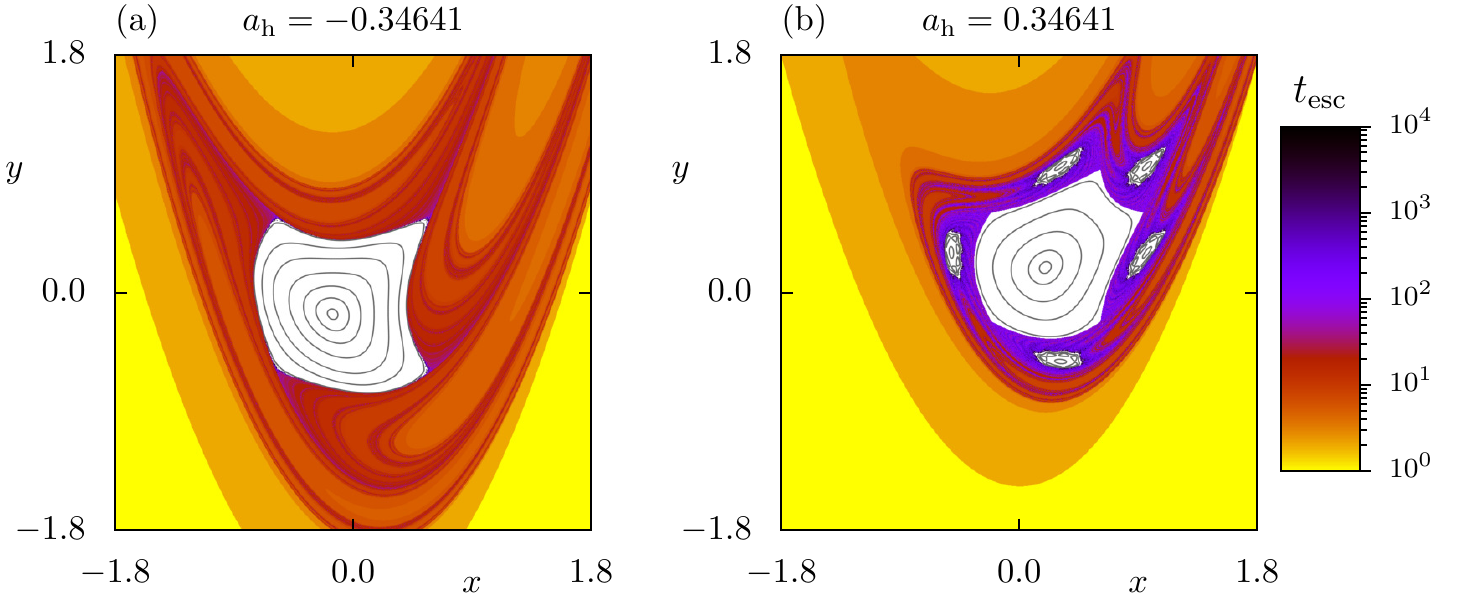}
\end{center}
\caption{Escape time plot for the H\'enon map for $\ahenon=-0.34641$ (left)
         and $\ahenon=0.34641$ (right).
         Some regular tori are shown as grey curves.}
\label{fig:henon-map}
\end{figure}

For the two-dimensional case, the map can be transformed
by an affine coordinate change to the form
\beq{HenonMap}
	H(x,y) =  (-y + \ahenon + x^2, x)
\eeq
with a single parameter $\ahenon$.
For $\ahenon < 1$ this map has two fixed points $(x,y) = (x^*_\pm, x^*_\pm)$, with
\begin{equation}
   x^*_\pm = 1 \pm \sqrt{1 - \ahenon}.
\end{equation}
These fixed points are created in a
saddle-center bifurcation when $\ahenon=1$ at $(x^*, y^*) = (1, 1)$.
When $\ahenon < 1$ the fixed point $(x^*_+,x^*_+)$ is hyperbolic.
When  $-3 < \ahenon < 1$, the negative branch is linearly stable (elliptic),
and the map is conjugate
to the map introduced by H\'enon \cite{Henon69}.
At $\ahenon=-3$ the lower fixed point undergoes a
period-doubling bifurcation, and
below this parameter value both fixed points are unstable.
It has been shown that there are no bounded orbits outside the square
$S = \{ |x| \le x^*_+, |y| \le x^*_+\}$ \cite{Devaney79}.
Figure~\ref{fig:henon-map} illustrates the dynamics
for two different parameters using an escape time plot:
for a grid of initial points the time needed to
leave the square $S$ is encoded in color.
Points that stay inside $S$ for $10^4$ iterations
are shown in white. The elliptic fixed point is
surrounded by invariant tori, as predicted by the
Kolmogorov-Arnold-Moser (KAM) theorem.
Some of these tori are shown as grey curves.
In the right panel, one can also see a period-five orbit
and some of the surrounding invariant circles that form
a resonance zone.

%%%%%%
%%%%%%%

By applying a similar coordinate change, Moser showed in \cite{Moser94}
that the quadratic symplectic map in $\bR^4$ can generically be written
in an analogous normal form. Transforming this slightly by shifting
the coordinates and parameters \cite{BaeMei2018:p}, this map can be written as
\beq{MoserMap}
   (\xi', \eta') = M(\xi,\eta )
                 = (\xi + C^{-T}(-\eta + C\xi + \nabla U(\xi)) , C\xi ).
\eeq
Here $\xi\equiv(\xi_1, \xi_2) \in \bR^2$, $\eta\equiv(\eta_1, \eta_2) \in \bR^2$
are canonically conjugate coordinates and momenta, respectively, and
\begin{align}
      C &= \Biggl(\begin{matrix} \alpha & \beta \\ \gamma & \delta
                 \end{matrix}\Biggr),
       \label{eq:CMatrix} \\
      U &= a\xi_1 + b\xi_2 + \tfrac12 c \xi_1^2+ \eps_2 \xi_1^3
            + \xi_1 \xi_2^2 .
            \label{eq:UPotential}
\end{align}
The map has two discrete parameters,
$\eps_1 \equiv \det(C) = \alpha \delta -\beta\gamma\  = \pm 1$,
and $\eps_2 \equiv \pm1$ or $0$.
The remaining six parameters are free
and are conveniently grouped into $(a, b, c)$,
determining the location of the fixed points (together with $\eps_2$)
and $(\alpha, \delta, \mu)$, determining (together with $\eps_1$)
the stability properties.
Here we define
\beq{muDefine}
	\mu = \beta + \gamma
\eeq
from which the off-diagonal elements of $C$ follow by
\bsplit{betagamma}
	\beta, \gamma &= \tfrac12 \mu \pm \sqrt{ \eps_1 -\alpha\delta +\mu^2/4} .\\
\esplit
As shown in \cite{BaeMei2018:p} the choice of sign for $\beta$
and $\gamma$ is unimportant.
Note that \Eq{betagamma} has real solutions only when
$\mu^2 \ge 4(\alpha\delta - \eps_1)$, and that $C$ is symmetric,
$\beta=\gamma=\mu/2$, only at the lower bound of this inequality.

%%%%%%%%%%%%%%%%%%%%%%%%%%%%%%%%%%%%%%%%%%%%%%%%%%%%%%%%%%%%%%%%%%%%%%%%%%%%%
\section{The quadfurcation}
%%%%%%%%%%%%%%%%%%%%%%%%%%%%%%%%%%%%%%%%%%%%%%%%%%%%%%%%%%%%%%%%%%%%%%%%%%%%%

The fixed points $(\xi^*, \eta^*)$ of the map \Eq{MoserMap}
correspond to critical points of the cubic polynomial \Eq{UPotential}. Thus
the coordinates $\xi^*$ satisfy
\beq{ShiftedFP}
	0 = \nabla U(\xi^*) = \Biggl(\begin{matrix}
						a +  c \xi_1^* + 3 \eps_2 \xi_1^{*2} + \xi_2^{*2} \\
						b + 2\xi^*_1 \xi^*_2
                                     \end{matrix}\Biggr) .
\eeq
The momenta are then given by $\eta^* = C\xi^*$.
The case $a=b=c=0$ is an organizing center for the solutions of \Eq{ShiftedFP}.
In this case the second component immediately implies that either $\xi_1^* = 0$
or $\xi_2^* = 0$. and then, whenever $\eps_2 \neq 0$, the first implies that both
$\xi_1^* = \xi_2^* = 0$. We call this the \textit{quadfurcation point}. Since
the matrix elements $(\alpha,\delta, \mu)$ are still free, this occurs
on a codimension-three surface in the six-dimensional parameter space.
There are at most four fixed points except when $\eps_2 = 0$, which has
a line of fixed points when $a=b=c=0$.
For simplicity we will assume in this paper that $\eps_2 \neq 0$ (the
case $\eps_2 = 0$ is discussed in \cite{BaeMei2018:p}).

More generally if $b \neq 0$ then
\Eq{ShiftedFP} implies that $\xi_1^* \neq 0$, so
$$
	\xi_2^* = -\frac{b}{2\xi_1^*}  .
$$
Substituting into the first component of \Eq{ShiftedFP}
then shows that $\xi_1^*$ must be a root of the scalar polynomial
\beq{PofV}
	P(v) = 3\eps_2v^4 + c v^3 + a v^2 + \tfrac14 b^2 .
\eeq
When $\eps_2 \neq 0$ this polynomial is quartic, so there are at most four
roots. Since the linear term vanishes, there is exactly one only at the quadfurcation point
$a=b=c=0$.

There are various regions in the $(a, b, c)$ parameter space
that have different numbers of fixed points.
As we discuss below, of particular interest are paths through the quadfurcation point
along which four new solutions are created out of nowhere.

%%%%%
\begin{figure}[b]
\begin{center}
\includegraphics{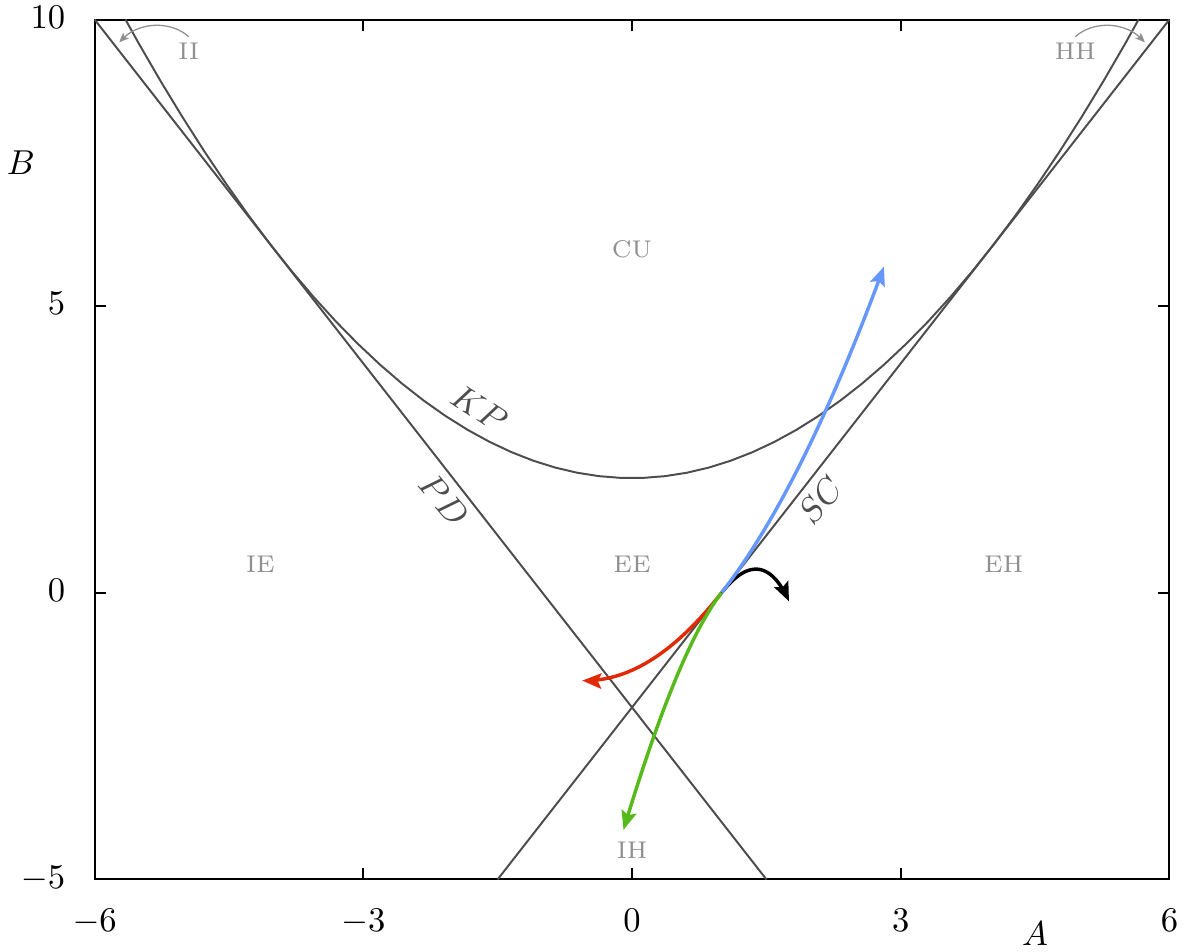}
\end{center}
\caption{ Stability parameters of the four fixed points in the $(A,B)$ plane created in the transition,
$\emptyset \to$ 2 EE + 2 EH for $\Delta <0$ along
the path \eqref{eq:quadfurcation-path}
with $(\alpha, \mu, \delta) = (1, 1, 0.5)$
and $\eps_1 = \eps_2 = 1$.
Since $|\beta-\gamma| = \sqrt{3}$, \Eq{ABQuad} implies that the quadrupling occurs at
$(A^\txtQ,B^\txtQ) = (1, 0)$.
}\label{fig:quadfurcation-AB}
\end{figure}
%%%%%

The stability of a fixed point is  characterized by
Broucke's parameters \cite{Broucke69, Howard87}
$A = \tr{DM}$ and
$B = \tfrac12 \left( (\tr{DM})^2-\tr{DM^2}\right)$,
where $DM$ is the linearized map at the fixed point.
There are seven stability regions in the $(A, B)$-plane
bounded  by the saddle-center ($SC$) line
\beq{SCLine}
	SC = B-2A+2 = 0,
\eeq
where there is a double eigenvalue $\lambda_1=\lambda_2 = 1$,
the period-doubling ($PD$) line
\beq{PDLine}
	PD = B+2A+2 = 0,
\eeq
where there is a double eigenvalue $\lambda_1 = \lambda_2 = -1$,
and the Krein parabola ($KP$)
\beq{KreinBifs}
	KP = B-A^2/4-2 = 0 ,
\eeq
where there are two sets of double eigenvalues on the unit circle or real line.
These regions are shown in \Fig{quadfurcation-AB}.
The stability type of a fixed point is labeled
by combinations of of E (elliptic), H (hyperbolic), and
I (inverse hyperbolic). For example EH denotes a point
with one elliptic and one hyperbolic pair of eigenvalues.
The seventh case, denoted CU (complex unstable), occurs in the region $KP>0$
and corresponds to a Krein quartet of complex eigenvalues.

Figure~\ref{fig:quadfurcation-AB}
shows a quadfurcation that
initially creates two EE and two EH fixed points.
The curves trace out the stability of the four fixed points
along the parameter path
\beq{quadfurcation-path}
  (a,b,c) = \Delta(1.0, 0.25, 0.5)
\eeq
with $\Delta\in[-0.5, 0]$, and the remaining parameters given in the caption.
For this matrix, the four fixed points
are born on the $SC$ line where the eigenvalues are $1$, $1$, and $\tfrac12(-1\pm \sqrt{3}\ui)$,
and emerge along paths that are tangent to this line.
This tangency holds generically for curves in parameter space that go through the quadfurcation point.
The EE point indicated by the red curve in the figure eventually loses stability
by period-doubling (for $\Delta = -0.31745$), becoming IE,
and that shown in blue loses stability by a Krein bifurcation
(for $\Delta = -0.21536$), becoming CU.

%%%%%%%%
\begin{table}[t]
\begin{center}
\begin{tabular}{c|c|c||c|c}
$(A^\txtQ, B^\txtQ)$  & \multicolumn{2}{c}{Condition}    &  \multicolumn{2}{c}{Fixed Points and Stability} \\
     on $SC$    &   $\eps_1$ & $|\beta-\gamma|$   &  $\eps_2=1$, $a < -\sqrt{3}|b|$  & $\eps_2=-1$, $a$ or $b\neq0$  \\\hline
$ > (4, 6)$        & $-1 $  & $\neq 0$            &  2 EH $+$ 2 HH & 2 HH  \\
$ < (4, 6)$        & $1$    &  $<2$           &  2 EE $+$ 2 EH & 2 EH  \\
$ = (0,-2)$		   & $1$    & $2$ & IE $+$ EE $+$ IH $+$ EH & IH $+$ EH \\
$ < (0, -2)$       & $1$,   &$> 2$ &  2 IE $+$ 2 IH & 2 IH   \\
\end{tabular}
\end{center}
\caption{Location of the quadfurcation along the $SC$ line (column one) and
stabilities of the created fixed points (columns four and five).
These stabilities are valid for a path of the form
$(a,b,c) = \Delta(a^*,b^*,c^*)$ which has a quadfurcation at $\Delta = 0$.
For $\eps_2 = 1$, four fixed points are created as
$a$ becomes negative provide that $a < -\sqrt{3}|b|$.
When $\eps_2=-1$, two fixed points exist whenever $a$ or $b \neq 0$; these collide at $\Delta = 0$.
The special case $\beta = \gamma$ is not shown here.
}
\label{tbl:overview}
\end{table}
%%%%%%%%

The quadfurcation occurs on the $SC$ line in \Fig{quadfurcation-AB}, and this is true more generally.
Indeed when $a=b=c=0$, then
\bsplit{ABQuad}
	A^\txtQ &= 4-\eps_1(\beta-\gamma)^2 ,\\
	B^\txtQ &= 6 -2\eps_1(\beta-\gamma)^2 ,
\esplit
which is on the line \Eq{SCLine}. Thus the position
along the line depends primarily upon $\beta-\gamma$, the asymmetric part of $C$ \Eq{CMatrix}.
When $C$ is symmetric, the quadfurcation occurs at $(A^\txtQ,B^\txtQ) = (4,6)$,
which corresponds to a quartet of unit eigenvalues. If $C$
is asymmetric and $\eps_1 =1$, it occurs below this point. The quadfurcation
then creates one pair of fixed points below $SC = 0$, of type EH or IH,
and one above $SC = 0$, (initially) of type EE or IE.  When $\eps_1 = -1$ the
quadfurcation occurs above $(4,6)$ on the $SC$ line and there will be two fixed points (initially)
of type HH and two of type EH.
A quadfurcation can directly lead to CU fixed points only when $C$ is symmetric.
However, when $C$ is asymmetric
then it is possible for an EE or an HH fixed point to undergo a Krein bifurcation
at some parameter value after it is created,
recall the blue curve in \Fig{quadfurcation-AB}.
The different stability combinations that arise from quadfurcations
along lines in $(a,b,c)$ space, like that in \Eq{quadfurcation-path},
are summarized in Tab.~\ref{tbl:overview}, see \cite{BaeMei2018:p} for details.

%%%%%%%%%%%%%%%%%%%%%%%%%%%%%%%%%%%%%%%%%%%%%%%%%%%%%%%%%%%%%%%%%%%%%%%%%%%%%
\section{Geometry --- elliptic bubbles}
%%%%%%%%%%%%%%%%%%%%%%%%%%%%%%%%%%%%%%%%%%%%%%%%%%%%%%%%%%%%%%%%%%%%%%%%%%%%%

Of particular interest are those parameters of the Moser
map that lead to doubly-elliptic (EE) fixed points.
As expected from Kolmogorov-Arnold-Moser theory \cite{BroSev2010},
when the twist is nondegenerate such fixed points should be surrounded by a
Cantor family of two-tori on which the dynamics is conjugate to incommensurate
rotation. The result that shows this for the \twoD{} case is Moser's twist theorem \cite{Moser62}.
It implies that the density of these tori approaches one as they limit on an elliptic point
provided that the linearized frequency is not in a low-order resonance.
We have not been able to find a statement of a similar result for higher-dimensional maps,
though results along these lines have been proven for elliptic
equilibria of Hamiltonian flows \cite{DelGut1996, Eliasson13}.

Motion on these invariant tori is, of course, bounded; but just
as for the H\'enon map, most orbits of the Moser map are unbounded.
Indeed one can show that when $\eps_2 \neq 0$, there exists a radius $\kappa$
such that if two successive points fall outside the ball,
$\|\xi_t\|, \|\xi_{t+1} \| > \kappa$,
then the orbit is unbounded (An upper bound for $\kappa$ is
obtained in \cite{BaeMei2018:p}).
One way to display the transition between
bounded and unbounded motion is an escape time plot,
similar to that for the H\'enon map in Fig.~\ref{fig:henon-map}:
initial points
are iterated until they leave the ball of radius $\kappa$,
and the required time to escape is encoded in color.
To visualize this, we consider a grid of initial conditions on a \twoD{} plane
in phase space, and plot the escape time for each point using a color scale.

\begin{figure}
\includegraphics{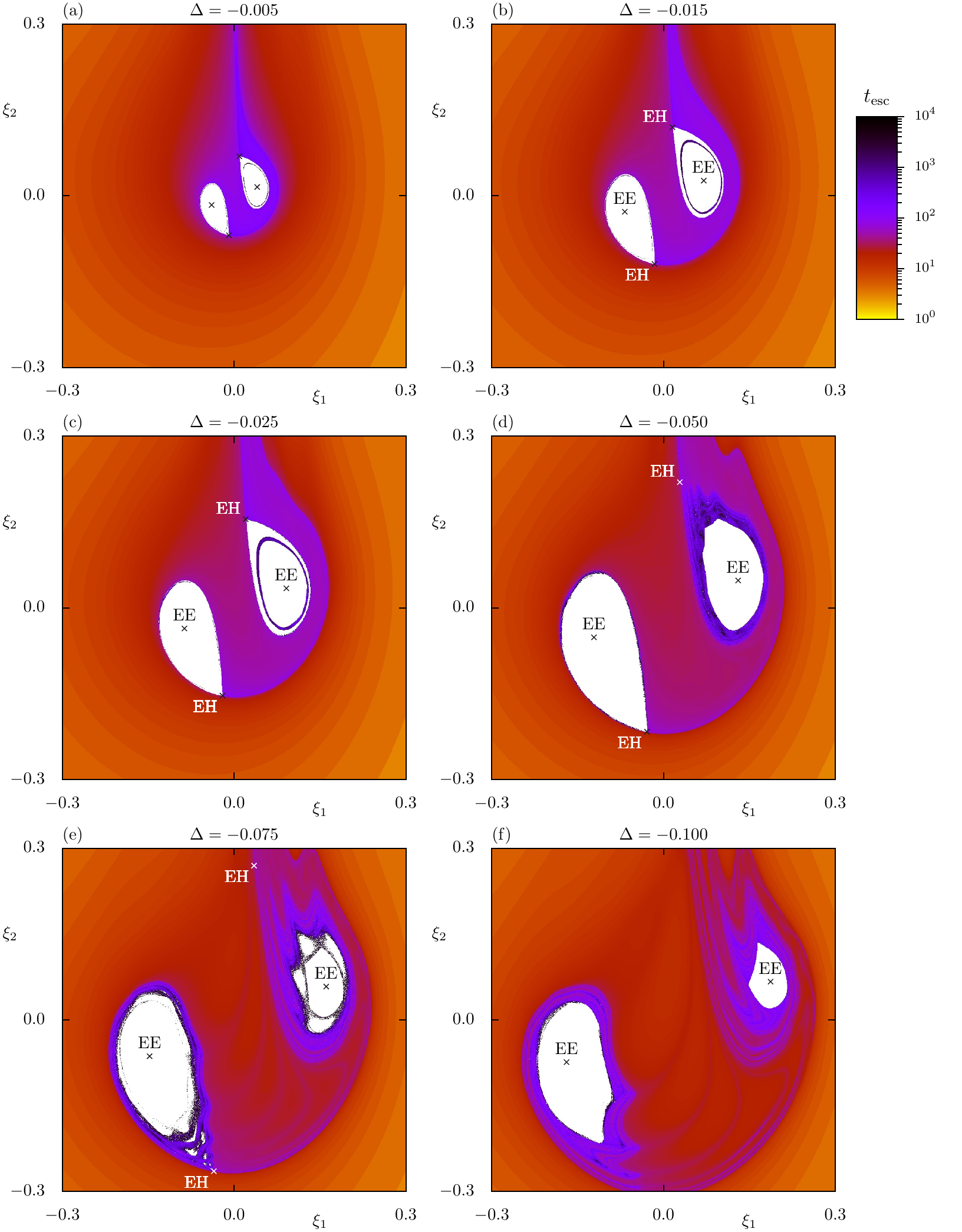}
\caption{Escape time plots  with the parameters of \Eq{quadfurcation-path} for
six values of $\Delta$,
$(\alpha, \mu,\delta) = (1,1,0.5)$, and $\eps_1=\eps_2 = 1$. Points on a $3000\times 3000$ grid
of initial conditions are iterated up to $t = 10^4$, and the escape time
is encoded in the color scale shown at the top right.
Points that do not escape are colored white.}
\label{fig:escape-plots}
\end{figure}

%%%%%%
\begin{figure}[t]
\includegraphics{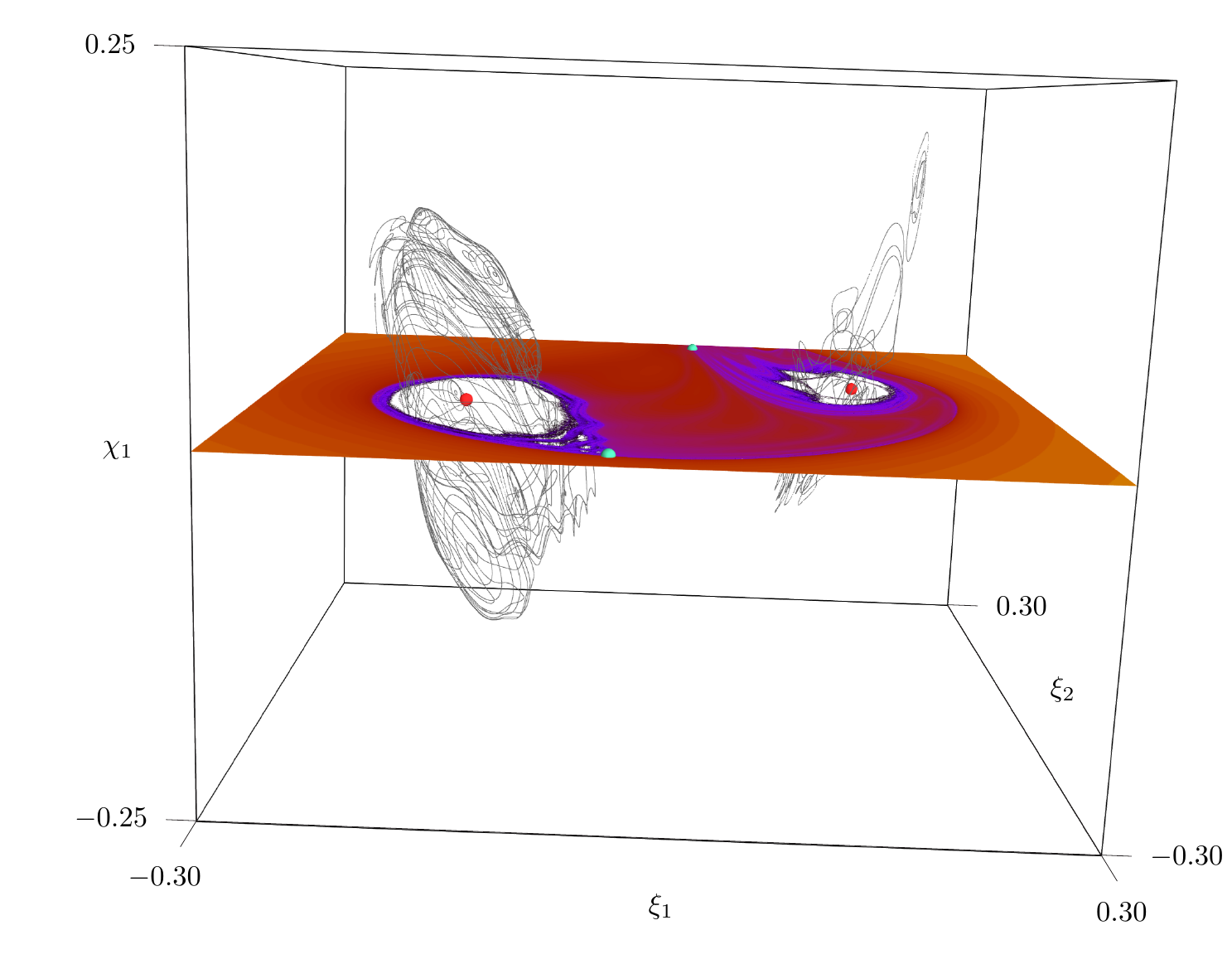}
\caption{3D phase space slice for the Moser map
         and corresponding escape time plot in the $\eta = C\xi$ plane.
         The parameters are the same as \Fig{escape-plots}(e) with
         $\Delta = -0.075$.
         Shown are several selected regular tori (grey curves)
         in the neighborhood of the EE fixed points.
         Each torus is represented by $10^4$ points that fall in the
         slice \Eq{slice} with $\epsilon=10^{-6}$.
         The four fixed points are shown as small spheres:
         EE (red), EH (green).
         The coloring of the escape times is the
         same as in \Fig{escape-plots}.
         \movierefall
         }
         \label{fig:quadfurcation-3dpss}
\end{figure}
%%%%%%%%

Figure~\ref{fig:escape-plots} shows escape time plots
for the quadfurcation of Fig.~\ref{fig:quadfurcation-AB}
for a sequence of $\Delta$ values along the path
\Eq{quadfurcation-path} as $\Delta$ varies, with the matrix $C$ held fixed.
Here we use the \twoD{} plane of initial conditions  $(\xi,\eta) = (\xi, C\xi)$
that contains all four fixed points of \Eq{MoserMap}. These are shown in the figures by $\times$ symbols.
After the quadfurcation, the regions near the four fixed points that are ``bounded"
are colored white: these correspond to orbits with numerically computed escape times larger than $10^4$.
The bounded orbits appear to be primarily associated with EE-EH pairs, each
leading to a structure reminiscent of that
of the \twoD{} H\'enon map shown in Fig.~\ref{fig:henon-map}, though now this structure
is seen in the two-plane $\eta = C\xi$ instead of the full phase space.
The elliptic-elliptic fixed points are surrounded by
regions of predominantly regular motion (see below).
One interesting feature, not possible for a \twoD{} map,
is that even ``within'' a region of regular
motion one finds orbits that can escape;
for example, there is a ring of escaping orbits near the right EE
fixed point in \ref{fig:escape-plots}(c).
Indeed, a linearly stable, elliptic fixed point can be unstable due
to Arnold's transition chain mechanism \cite{Arnold64, Lochak99,DelHug2011,Dum2014}.
Numerically one sees that resonances between the oscillation frequencies of surrounding tori
lead to chaotic zones and drift along resonant channels leads to additional escape routes.

To help visualize the geometry of regular tori
we can also use a \threeD{} phase space slice \cite{RicLanBaeKet2014}.
For this we define new coordinates $(\xi,\chi)$, with $\chi = \eta-C\xi$,
so that the fixed points lie in the \twoD{} plane $\chi = 0$. In order to capture
points on trajectories in the neighborhood of the fixed points, we define the slice
\beq{slice}
    \Gamma_{\epsilon} =
     \left\{ (\xi_1, \xi_2, \chi_1, \chi_2) \; \left| \rule{0pt}{2 ex} \;
         |\chi_2| \le \epsilon \right. \right\} ,
\eeq
which is a slightly thickening of the \threeD{} plane $\chi_2 = 0$.
Whenever the points of an orbit, given by a sequence of points
$(\xi_1, \xi_2, \chi_1, \chi_2)$, lie in this slice,
the three nontrivial coordinates $(\xi_1, \xi_2, \chi_1)$ are displayed in a \threeD{} plot.
For further examples and detailed discussion see
\cite{RicLanBaeKet2014,LanRicOnkBaeKet2014,OnkLanKetBae2016,FirLanKetBae2018}.

%%%%%%%%
\begin{figure}
\includegraphics{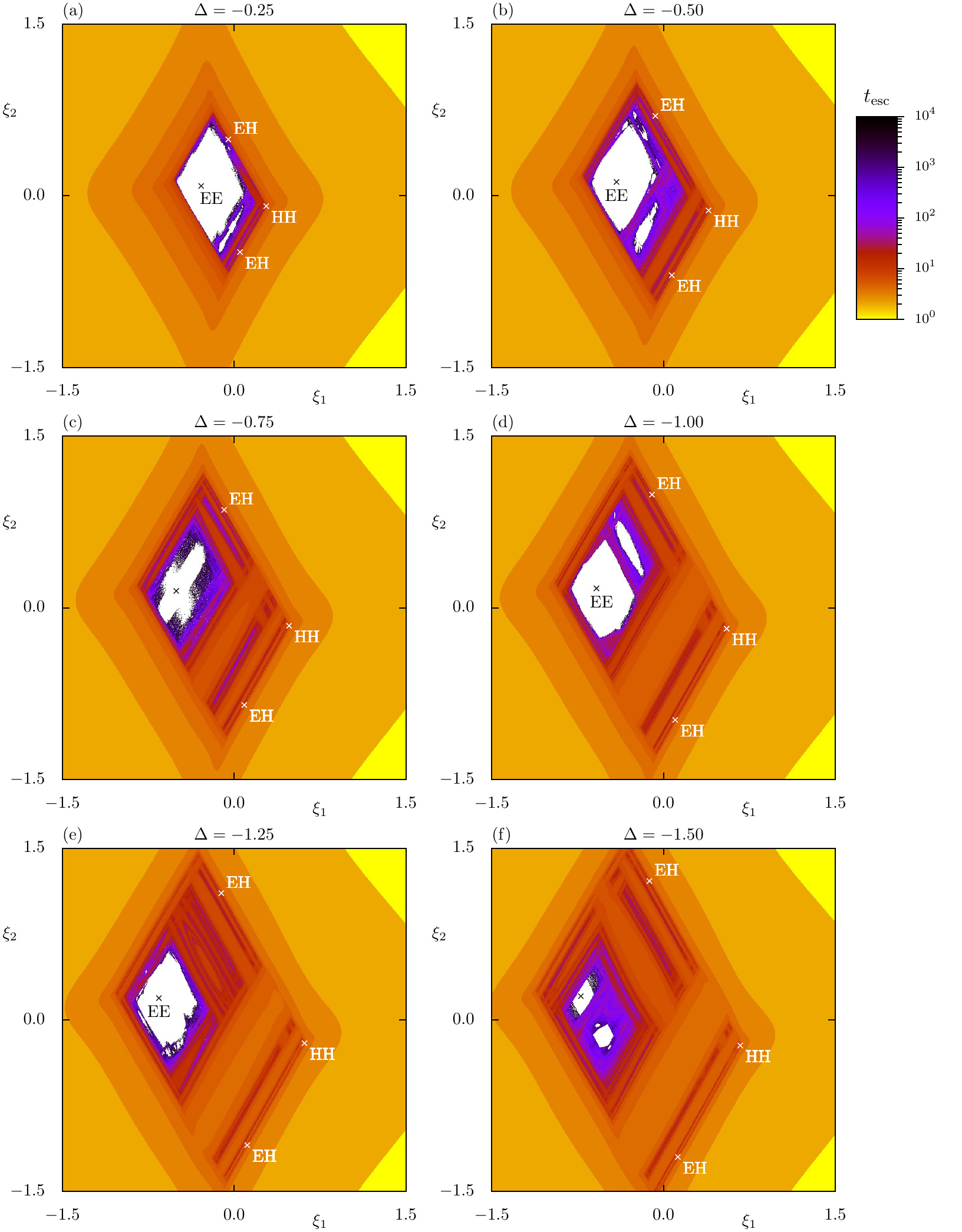}

\caption{Escape time plots for six values of $\Delta$ with $(a, b, c) = \Delta (1, -0.2, -0.1)$,
$(\alpha, \mu,\delta) = (\sqrt{3}, 0, 1/\sqrt{3})$,
and $\eps_1=\eps_2 = 1$.
Points on a $3000\times 3000$ grid
of initial conditions are iterated up to $t = 10^4$, and the escape time is encoded in the color scale shown at the top right.
Points that do not escape up to $t=10^4$ are colored white.}
\label{fig:escape-plots2}
\end{figure}
%%%%%%%

Figure~\ref{fig:quadfurcation-3dpss}
shows a \threeD{} phase space slice plot combined with
an escape time plot for a grid of initial conditions
in the \twoD{}-plane $\chi = 0$,
for the quadfurcation of Fig.~\ref{fig:quadfurcation-AB}
when $\Delta=-0.075$, which corresponds to  Fig.~\ref{fig:escape-plots}(e).
The elliptic-elliptic fixed points are surrounded by
regions of predominantly regular motion.
The intersections with the slice of a number of regular \twoD{} tori
are shown as grey curves.
Various resonances lead to gaps of different sizes, and these
are particularly visible in the tori that surround and approach
the ``left" EH fixed point (that with $\xi_1 < 0$).
Between the tori one has chaotic motion
in resonance channels that eventually may escape from the
neigbhorhood of the EE fixed points.
While the extent of each region with regular tori
in the $(\xi_1, \xi_2)$--plane is approximately limited
by the corresponding EH fixed point,
the region containing tori seems to extend farther from the
EE points in the $\chi_1$ direction.

\pagebreak

\begin{figure}[t]
\includegraphics{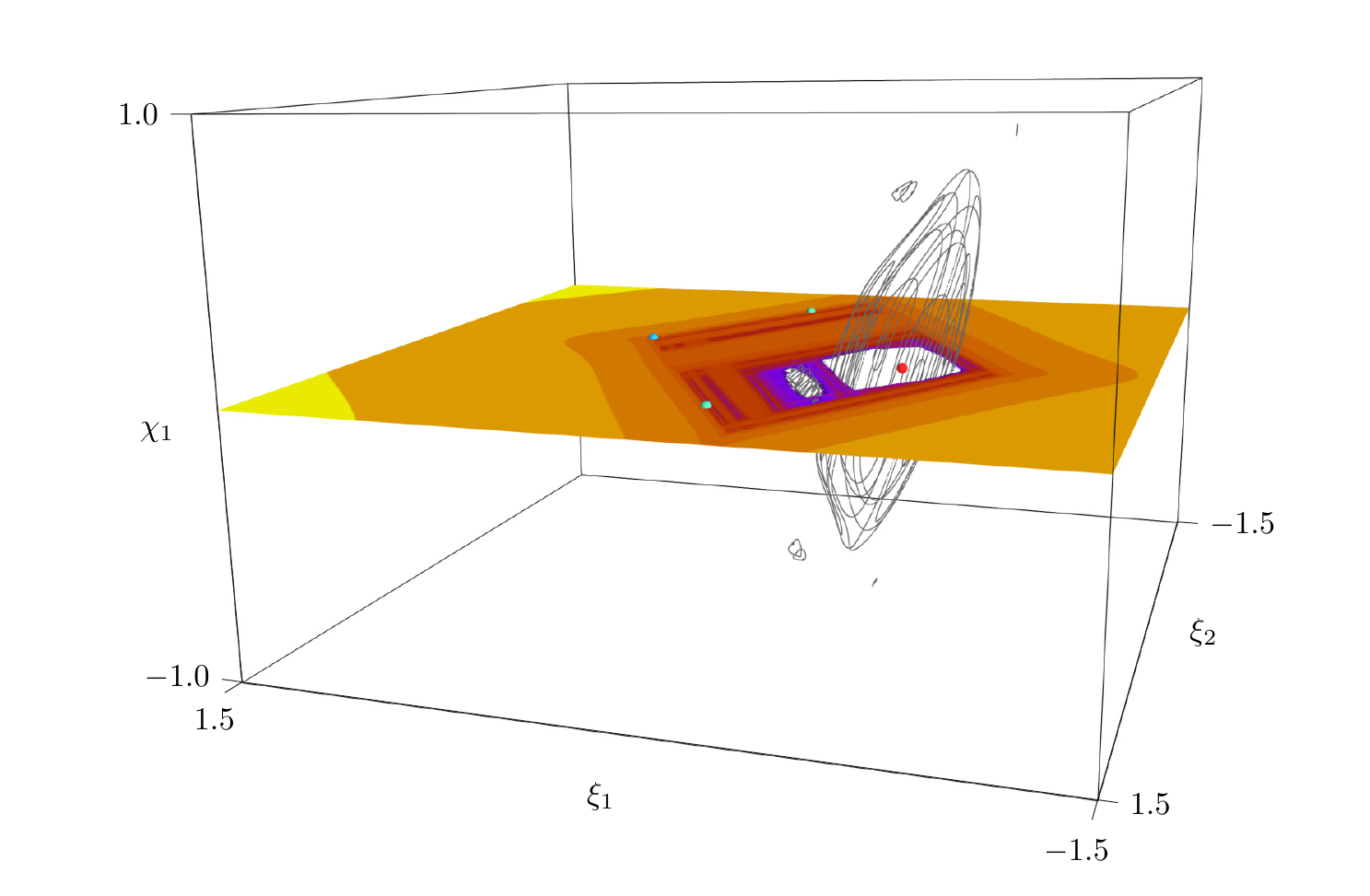}
\caption{3D phase space slice for the weakly coupled Moser map
         and corresponding escape time plot in the $\eta = C\xi$ plane.
         Parameters are the same as \Fig{escape-plots2}(d) with $\Delta = -1$.
         Shown are several selected regular tori (grey curves)
         near the EE fixed point.
         Each torus is represented by $10^4$ points in the
         slice \Eq{slice} with $\epsilon=10^{-6}$.
         The four fixed points are shown as small spheres:
         EE (red), EH (green), and HH (blue).
         The coloring of the escape times is the
         same as in \Fig{escape-plots2}.
         \movierefall
         }
         \label{fig:quadfurcation-3dpss-2}
\end{figure}

In contrast, there are two special cases for which Moser's map reduces to a pair of uncoupled H\'enon maps \Eq{HenonMap}. For example when $\eps_1 = +1$, $c = 0$, and
\[
	C = \begin{pmatrix} \sqrt{3} & 0 \\ 0 &\tfrac1{\sqrt{3}} \end{pmatrix} ,
\]
then the Moser map is equivalent (after a coordinate change) to two H\'enon maps with $\ahenOne = 1 + a - \sqrt{3} b$ and $\ahenTwo =  1+ a + \sqrt{3}b$. In this case the quadfurcation, at $a=b=0$ corresponds to simultaneous saddle-center bifurcations of the individual H\'enon maps, creating four fixed points of types EE, EH, EH and HH. Note that this quadfurcation occurs at the point $(A,B) = (4,6)$ where there is a quadruplet of unit eigenvalues. When the parameter $c$ is nonzero, these maps become coupled, but when $c$ is small the stability of the four fixed points is unchanged.
As an illustration a sequence of escape time plots for
the path $(a, b, c) = \Delta (1, -0.2, -0.1)$
through this quadfurcation is shown in \Fig{escape-plots2}.
The coupling is rather weak so that the
product structure of the two \twoD{} H\'enon maps,
as shown in Fig.~\ref{fig:henon-map}, corresponding to $(a, b) = (-1, 0.2)$
and $c=0$, is still visible.
Specifically, the dynamics along line connecting the EE
fixed point to the lower EH fixed point
in \Fig{escape-plots2}(d) is similar to that on the diagonal $y=x$ of the \twoD{} map
shown in \Fig{henon-map}(a), and would be identical if $c$ were $0$. Similarly
the dynamics along the line connecting
the EE fixed point to the upper EH fixed point
is governed by that shown in Fig.~\ref{fig:henon-map}(b)
along its diagonal, $y=x$.
In particular the splitting of the white region of non-escaping
points is caused by the stochastic motion around the period 5-island
in the \twoD{} map.
This is also reflected in the structure of regular
tori displayed in the \threeD{} phase space slice plot
shown in Fig.~\ref{fig:quadfurcation-3dpss-2},
where the resonance leads to several small regions with regular tori.
Again the EE fixed point is surrounded by many regular tori
shown as grey curves.

%%%%%%%%%%%%%%%%%%%%%%%%%%%%%%%%%%%%%%%%%%%%%%%%%%%%%%%%%%%%%%%%%%%%%%%%%%%%%
\section{Some implications and discussion}
%%%%%%%%%%%%%%%%%%%%%%%%%%%%%%%%%%%%%%%%%%%%%%%%%%%%%%%%%%%%%%%%%%%%%%%%%%%%%
\enlargethispage{4ex}

Our investigation of Moser's quadratic symplectic
map in $\bR^4$ reveals that for the quadfurcation,
in which four fixed points are created from none,
there are several distinct possibilities;
one is the expected combination of two uncoupled
H\'enon maps, leading to one EE, two EH, and one HH fixed points,
illustrated in Figs.~\ref{fig:escape-plots2}
and \ref{fig:quadfurcation-3dpss-2}.
This case also describes accelerator mode islands
\cite{ChiIzr1973,KooMei1990} that are born for the
\fourD{} standard map (Froeschl\'e's map \cite{Fro1971}) when
the coupling is small. Here the local dynamics reduces to a coupled version of a pair of
H\'enon maps \cite{BaeMei2018:p}.

Perhaps more surprising is the quadfurcation that
creates two pairs of EE and EH fixed points,
as illustrated in Figs.~\ref{fig:escape-plots}
and \ref{fig:quadfurcation-3dpss},
which is only possible when different degrees of freedom are coupled.

It will be interesting in future research to do
detailed studies of accelerator mode islands in the fully coupled \fourD{} case,
to study in more detail the stickiness and survival time statistics for
Moser's map in $\bR^4$, and to investigate
what analogues of quadfurcation occur in higher dimensions.

\section*{Acknowledgments}
JDM acknowledges support from the U.S. National Science Foundation under grant DMS-1812481,
and as Dresden Senior Fellow at the Technische Universit\"at Dresden.
AB acknowledges support by the Deutsche Forschungsgemeinschaft under grant KE~537/6--1. The visualizations of the \threeD{} phase space slices were created using \textsc{Mayavi}~\cite{RamVar2011}.

%%%%%%%%%%%%%%%%%%%%%%%%%%%%%%%%%%%%%%%%%%%%%%%%%%%%%%%%%%%%%%%%%%%%%%%%%%%%%
%\section*{References}
%%%%%%%%%%%%%%%%%%%%%%%%%%%%%%%%%%%%%%%%%%%%%%%%%%%%%%%%%%%%%%%%%%%%%%%%%%%%%

\renewcommand{\url}[1]{\href{#1}{\texttt{#1}}}

\end{document}